\documentclass{article}

\usepackage[english]{babel}

\usepackage[letterpaper,top=1.5cm,bottom=1.5cm,left=2.2cm,right=2.2cm,marginparwidth=1cm]{geometry}

\usepackage{amsmath}
\usepackage{graphicx}
\usepackage[colorlinks=true, allcolors=blue]{hyperref}
\usepackage{authblk}

\title{Preserving Nature's Ledger: Blockchains in Biodiversity Conservation}

\author[1]{Kostas Kryptos Chalkias}
\author[2]{Angelos Kostis}
\author[1,3]{Ali Alnuaimi}
\author[4,5]{Peter Knez}
\author[1]{John Naulty}
\author[6]{Allen Salmasi}
\author[1]{Ryan Servatius}
\author[7]{Rodrigo Veloso}
\affil[1]{Mysten Labs}
\affil[2]{Stanford University}
\affil[3]{Shafra}
\affil[4]{Nobel Sustainability Science Academy}
\affil[5]{InCapture Group}
\affil[6]{Veea}
\affil[7]{O.N.E. Amazon}

\begin{document}
\date{}
\maketitle

\begin{abstract}
    In the contemporary era, biodiversity conservation emerges as a paramount challenge, necessitating innovative approaches to monitoring, preserving, and enhancing the natural world. This paper explores the integration of blockchain technology in biodiversity conservation, offering a novel perspective on how digital resilience can be built within ecological contexts. Blockchain, with its decentralized and immutable ledger and tokenization affordances, presents a groundbreaking solution for the accurate monitoring and tracking of environmental assets, thereby addressing the critical need for transparency and trust in conservation efforts. Unlike previous more ``theoretical'' approaches, by addressing the research question of how blockchain supports digital resilience in biodiversity conservation, this study presents a grounded framework  that justifies which blockchain features are essential to decipher specific data contribution and data leveraging processes in an effort to protect our planet's biodiversity, while boosting potential economic benefits for all actors involved, from local farmers, to hardware vendors and artificial intelligence experts, to investors and regular users, volunteers and donors. 

\end{abstract}

\setcounter{page}{0}
\thispagestyle{empty}
\newpage
\section{Introduction}
Biodiversity conservation is like earth's life insurance policy. It's all about keeping the planet's amazing array of plants, animals, and ecosystems safe and healthy. This means managing habitats, protecting different species, and preserving genetic diversity. The ultimate aim is to ensure that the natural world can thrive and adapt, even as it faces challenges like climate change and human activity \cite{colwell09}; this is also a key factor for humanity's survival and future prosperity.

In the globalized financial scene, coexisting biodiversity with environmental changes and economic activities like farming and mining \cite{brooks2006global} can potentially be possible through sustainable practices and the use of modern technology such as artificial intelligence (AI), internet of things (IoT) and blockchains in an effort to minimize negative impacts on ecosystems and species \cite{ijerph16203847}. On the policy side, this can include using eco-friendly farming techniques that conserve soil and water, protect biodiversity-rich areas from mining activities, and implementing restoration programs to rehabilitate degraded habitats \cite{devarinti16}. On the technical front though the mere assertion of potential societal benefits of AI and blockchain technologies, without a thorough examination of their practical implications and limitations, has led to premature and overly enthusiastic or pessimistic conclusions. This work focuses on providing tangible claims and justified technological guidance on what Web3 features and tools are available today in order to provide an answer to one of the most common criticism of often characterizing blockchain as mere hype \cite{risius2017blockchain}.

Investigating how blockchain can allow biodiversity conservation, we address a need ``for the field to dwell on the societal implications of the technology'' \cite{beck2017blockchain}, its relevance to grand challenges \cite{avital2016jumping}, and what are the realistic primitives that can only be offered by such an infrastructure, rather than just theoretical claims. In doing so, we introduce a distinctive approach to building digital resilience \cite{boh2023building} not only for responding to immediate crises, such as earthquakes or pandemics, but also for addressing more pervasive environmental challenges, including threats to biodiversity. This approach involves leveraging the decentralized and transparent logic of underlying blockchains and highlighting that natural resources and ecosystems are more valuable alive as the involved species and entities can be assigned a tradable digital value \cite{knez24}.


An emerging body of literature has started paying attention to digital resilience \cite{sakurai2020resilience, floetgen2021introducing, park2023value}, defined as “the capabilities to absorb major shocks, adapt to disruptions caused by the shocks, and transform to a new stable state, where entities are more prepared to deal with major shocks.” \cite{boh2023building}. Research in that space has taken some important steps towards deepening our understanding of the role of digital technologies in facilitating responses to exogenous shocks \cite{heeks2019conceptualising,tim2023digital,liu2023understanding}. For instance, it has been reported that individuals, organizations, and ecosystems can build digital resilience capabilities through centralized IT investments \cite{park2023value} and by relying on data and digital resources for decision making during the shock  \cite{tremblay2023data}. Relatedly, physicians in China used digital platforms to maintain their services immediately after the COVID-19 outbreak and recover in the post-shock period, thereby both absorbing the shock and transforming to a new more adaptable state \cite{liu2023understanding}. 

Nonetheless, current literature on digital resilience has two main limitations. First, while there has been a focus on dealing with short-term shocks, little is known about how digital technologies aid in dealing with major and pervasive environmental crises, such as threats to biodiversity, that stretch over time and have disastrous implications for the humanity. Interestingly, recent data shows that we should classify biodiversity loss as a major global shock that negatively transforms life in our planet at alarming rates; experts estimate that the current rate of species loss is between 1,000 and 10,000 times faster than the natural rate of extinction \cite{wwf21}. Second, most of the attempts are essentially regional efforts and mostly centralized, despite voices pointing to blockchain as supporting the transition towards a sustainable global economy by transforming critical for the society sectors \cite{nguyen2016blockchain}. In fact, digital resilience research has overlooked the role of blockchain in producing solutions to major threats to biodiversity. This is surprising given that blockchain technology has been acknowledged not only as a game-changer for sustainability but also as a ``valuable enabler of economic and social transactions'' \cite{beck2017blockchain}. In addition, research notes that blockchain is  a revolutionary technology supporting trust and collaborations among organizations \cite{lumineau2021blockchain} while enabling fundamentally new forms of organizing rooted in collective action \cite{beck2018governance,ellinger2023skin}. Yet, how blockchain can allow for and boost collective action for biodiversity conservation remains an important knowledge gap. Surprisingly, only a handful papers highlight the importance of this technology regarding environmental issues \cite{chapron2017environment}, and most of the previous attempts fail to substantiate the theoretical claims and to focus on recently appeared novel primitives in the fields of blockchain compression, consensus, publicly verifiable audits, e-voting, cryptography, tokenization, privacy, cryptography, combining AI with IoT and Web3 technologies as one system and many more \cite{atlam2020review, jichalkias21, raikwar2019sok}. 
\section{Selective Blockchain Features}
A blockchain is a decentralized, transparent, and immutable public ledger that securely records transactions using advanced cryptography. It operates on a consensus mechanism, ensuring that no single entity can alter the recorded data unilaterally. This technology provides a secure platform for verifying ownership and transactions, fostering trust among participants through its inherent transparency and the collective agreement on any additions to the ledger. The importance of blockchains has led to a number of breakthrough innovations in the database, consensus, economics, tokenization, programming languages, security, and cryptography spaces, among others, which in turn attracted large investments from reputable funds and venture capitals.

Roughly since 2009, when the foundational launch of Bitcoin \cite{nakamoto2008bitcoin} introduced the decentralization concept, a number of specialized blockchain solutions emerged. Examples include the innovation of smart contracts with Ethereum \cite{wood2014ethereum}, the privacy-focused designs of Zcash and Monero \cite{akcora2022blockchain}, the enterprise-oriented solutions of Hyperledger and Corda \cite{monrat2020performance}, the scalability and performance advancements of blockchains like Solana, Sui, Aptos, Celestia and Avalanche, the efficiency improvements of Layer 2 technologies such as Arbitrum and zkSync and others \cite{puschmann2024taxonomy}. Together, these developments highlight the dynamic evolution of blockchain technology, driven by a continuous quest for improved performance, privacy, scalability, and utility.

Throughout this paper we will discuss selective key properties of blockchains \cite{raikwar2019sok} that have direct applications \cite{abou2019blockchain} to biodiversity conservation efforts. This is a major differentiator factor of our analysis against previous mostly theoretical works. These properties include:

\begin{itemize}
    \item \textbf{Decentralization}: Unlike traditional ledgers or databases that are controlled by a central authority (like a bank or government agency), a blockchain operates across a distributed network of computers. In short, no single entity can dominate transactions or shut down the network.
    This trait is crucial for international conservation efforts, where collaboration across borders and interests is necessary, while at the same time allows for seamless inclusion of more actors during the process.
        
    \item \textbf{Transparency and Immutability}: Blockchains offer transparency, enabling users to view and trace past transactions, building trust with verifiable actions. Additionally, confirmed and timestamped transactions are immutable, preventing alteration and ensuring reliable record-keeping. Both features are essential for the integrity of international  bio-conservation efforts, by demotivating malicious activity.

    \item \textbf{Smart Contracts}: Modern blockchains support smart contracts, which are self-executing contracts with the terms directly written into code. These contracts automatically enforce and execute the terms of agreements, providing a new level of automation and reducing the need for intermediaries \cite{zheng2020overview}. Importantly, smart contracts can encode a wide array of logic, making them versatile for various applications required for conservation policy making and governance such as managing donations, land registry, implementing  e-voting, auctions and marketplaces, among others.

    \item \textbf{Tokenization}: Tokenization transforms physical assets, rights, or values into digital tokens, streamlining asset transfers and enabling digital economies. It acts as a crucial incentive for developing a conservation economy, rewarding technology providers and users involved with tokenized natural assets \cite{un23}. A compelling concept is making each animal and plant more valuable alive and conserved within this tokenized ecosystem than through poaching or deforestation. A crucial point for this to work is providing incentives to local workers, such as farmers and miners, but also technology providers such as IoT, drone and network vendors. Blockchain smart contracts facilitate this by automating the distribution of incentives.

    \item \textbf{Oracles}: One of the most important features is allowing devices and committees to provide data from the external world to the blockchain, such as weather stations, drone pictures and IoT sensors. Ensuring data flows into and off a public immutable ledger enables a great set of applications from AI-based resolution of digital resilience (auto-react in wildfire, flood), to gamification and education via live non-censored updates.
        
        
        
    \item \textbf{Interoperability}: Blockchains enable information sharing across different systems, enhancing value by allowing cross-country, forest, and jurisdiction transactions and data exchanges, thus reducing the need for separate database integrations, which was previously a serious compatibility blocker for innovation.
        
    \item \textbf{Rapid Settlement and Cost Reduction}: Blockchain streamlines settlements, cutting time and costs; essential for swift funding in conservation crises. This efficiency benefits various sectors, like supply chain in emergencies where quick aiding and cost-effectiveness is essential for conservation initiatives.
        
    \item \textbf{Privacy}: Some blockchains allow technologies like zero-knowledge proofs to ensure transactions are both secure and private, blending transparency with confidentiality. This is crucial for individuals, such as farmers and IoT vendors, who require privacy for their data while contributing to shared prediction models.
       

\end{itemize}





\section{Towards a Framework of Blockchain for Digital Resilience }


The proposed framework presents a holistic, "blockchain features-driven" approach to biodiversity conservation, emphasizing the role of data collection, traceability, incentives, AI, IoT, and decentralised physical infrastructure networks (DePINs) \cite{ballandies2023taxonomy}. 

A cornerstone of the proposed framework is that the adoption of \textbf{\textit{tokenization}} strategies for biodiversity species and for IoT solutions, such as sensors, drones, and satellites to monitor and record data related to species and ecosystems, is crucial. Not only tokens can be created based on the logic of ``adopting a tree'', but also based on the logic of ``incentives to launch an IoT censor device''. In addition, blockchain offers possibilities for trading financial instruments \cite{HOWSON20191} such as green bonds and derivatives based on biodiversity data that IoT devices will collect about trees and other species \cite{oneamazon, un23}. 

\begin{figure}
    \centering
    \includegraphics[width=1\linewidth]{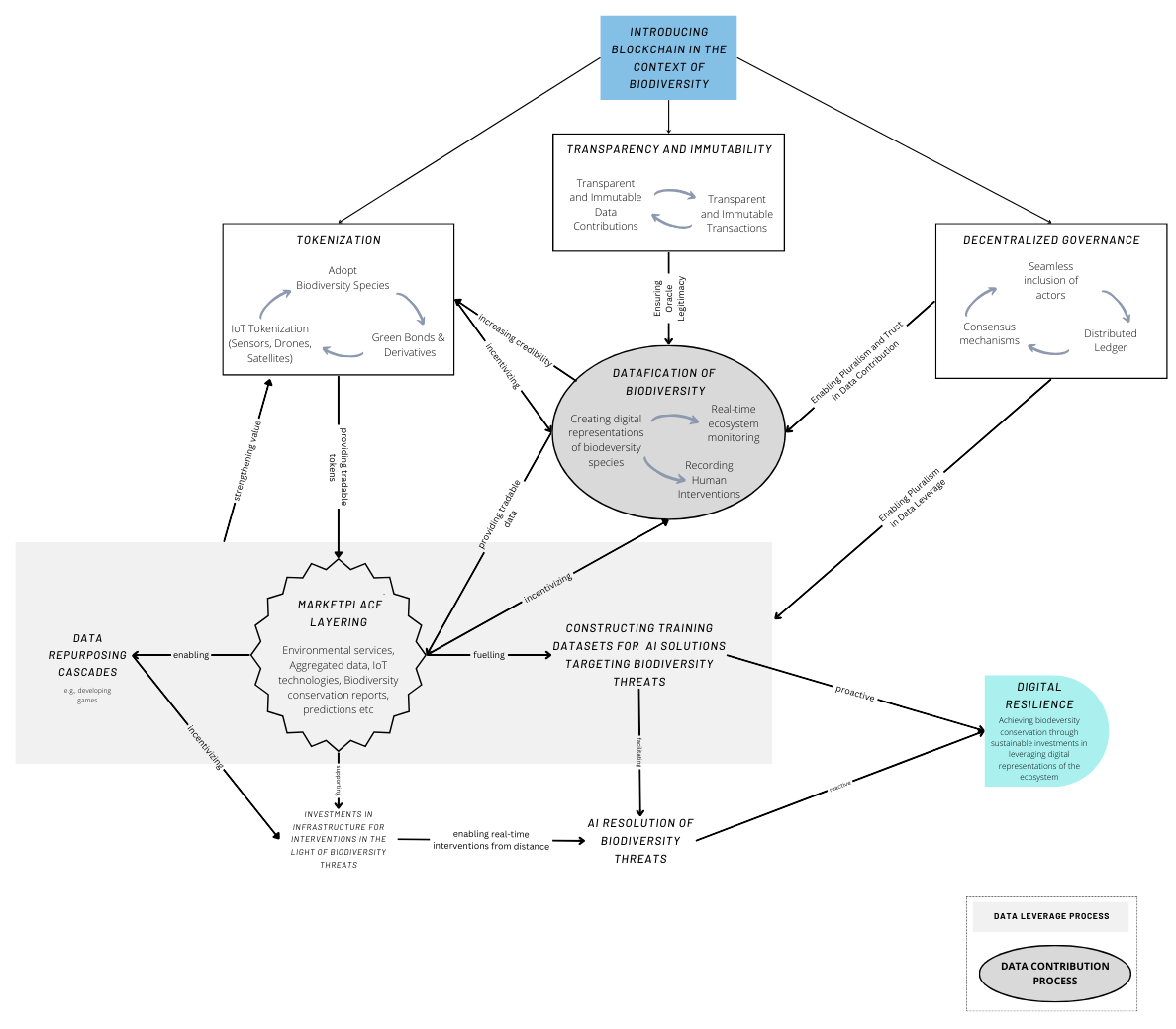}
    \caption{Blockchain and data processes enabling digital resilience within biodiversity conservation.}
    \label{fig:biodiversity-blockchain-framework}
\end{figure}

This, however, requires \textbf{\textit{datafication of biodiversity}}, a term we use to refer to the process of creating digital representations of various species in the ecosystem. As Fig.~\ref{fig:biodiversity-blockchain-framework} suggests, creating digital representations of biodiversity species and activities includes real-time ecosystem monitoring and recording human interventions that may improve or undermine biodiversity conservation. Here, it is important to note that the features of blockchain technology, and in particular the involved \textbf{\textit{transparency and immutability}}, safeguard the permanent presence of data contributions in a transparent way and build trust via \textbf{\textit{oracle}}s’ (contributors’) legitimacy. It is also highlighted that accumulating data and verifiably posting it in on-chain, requires specialized  common measurements for ecological environments, specifically forests, which primarily consist of LiDAR, Satellite Imagery, Drone Fleets, Field Measurements and Tamper-Proof Agents using advanced cryptography \cite{measure-monitor-redd-goetz, almashaqbeh2022sok}. Moreover, mesh networks \cite{veea} and internet-less transaction capabilities are essential for these systems to work during natural disasters and in remote environments, such as jungles, underwater and mountain terrains \cite{suiinternetless, musaddiq2022internet} .  


In addition, blockchains promote a \textbf{\textit{decentralized governance }}approach involving various actors and consensus mechanisms within a Web3 framework. Secure \textbf{\textit{smart contracts}} for e-voting, sealed-bid auctions and dynamic Non-Fungible Tokens (NFT) \cite{app11219931, nft} offer inclusivity and safeguard decision making transparency, even when participants' interests are not necessarily aligned.

While datafication refers to a data contribution process that is highly incentivized by the particular affordances of blockchains, our model shows that this technology also provides humanity with opportunities to leverage data in innovative ways. More specifically, we discuss the biodiversity\textbf{\textit{ data leveraging process}} as involving three key practices: marketplace layering, data repurposing cascades, and constructing training datasets for AI solutions targeting biodiversity threats.

\textbf{\textit{Marketplace layering}}. First, based the need to further expand data collection to cover as many species as possible from as many data sources as possible, an expansive and highly scalable biodiversity marketplace can be created. In particular, actors involved in data contribution or are interested in leveraging the collected data can create multiple layers within a marketplace that deals with biodiversity, where services, data, and reports can be traded. We propose that environmental services, aggregated biodiversity and IoT censor data, farming activities, gamification, educational content creation, and conservation reports can be part of this marketplace, all of which will facilitate the creation of new tokens and will further provide incentives to continue the datafication efforts.

\textbf{\textit{Data repurposing cascades}}. Second, we introduce the notion of “data repurposing cascades” to refer to the process of reusing collected biodiversity data for various purposes, enabling innovations and further developments that are not necessarily and directly connected to conservation efforts. An example here is the use of biodiversity data for constructing reports for magazines or even creating video-games and documentaries.

\textbf{\textit{Constructing training datasets for AI solutions targeting biodiversity threats.}} Third, our framework indicates that all collected data can be used to create AI training sets aimed at identifying and resolving threats to biodiversity. This is key because it can provide the possibility to engage with biodiversity threats in a proactive way based on predictions offered by the trained models or even in a reactive way where bushfires are for instance timely recognized and dealt with autonomous solutions, such as swarm intelligence drones that are in contact with sensors on the ground. Yet, this requires investments in infrastructure and various digital technologies and sensors as those are essential to enable real-time interventions to manage threats and shocks to biodiversity timely and effectively. Overall, on/off-chain AI can be deployed to analyze the data and proactively manage threats to biodiversity, while also potentially enabling reactive interventions from a distance.

Overall, the proposed model deciphers data contribution flows (see Fig.~\ref{fig:biodiversity-blockchain-framework}) through which blockchain technology incentivizes the creation and contribution of ecosystem's digital representations and facilitates the scalability and composability, via multiple independent actors being involved, of collected data types. It also underscores the critical role of blockchain in ensuring the integrity and security of data contributions and establishes oracles' legitimacy. Moreover, the model presents specific data leverage mechanisms as discussed above.

The results of leveraging data will hopefully help in biodiversity preservation via sustainable digital technology investments, thereby guaranteeing digital resilience and lasting protection of environmental ecosystems.


\bibliographystyle{apalike} 
\bibliography{bibtex}

\end{document}